\documentclass[twocolumn]{aastex631}
\received{x}
\revised{x}
\accepted{x}
\submitjournal{ApJL}
\usepackage{fancyvrb} 
\usepackage{breakurl}
\usepackage{multirow}
\usepackage{xcolor}
\VerbatimFootnotes    

\begin{document}

\newcommand{\vdag}{(v)^\dagger}
\newcommand\aastex{AAS\TeX}
\newcommand\latex{La\TeX}

\shorttitle{Neutrinos from TXS~0506+056}
\shortauthors{J.~Becker Tjus et al}

\title{Neurino Cadence of TXS~0506+056 Consistent with Supermassive Binary Origin}

\correspondingauthor{Julia Becker Tjus}
\email{julia.tjus@rub.de}

\author[0000-0002-1748-7367]{Julia Becker Tjus}
\affiliation{Theoretical Physics IV: Plasma-Astroparticle Physics, Faculty for Physics \& Astronomy, Ruhr University Bochum, 44780 Bochum, Germany}
\affiliation{Ruhr Astroparticle And Plasma Physics Center (RAPP Center), Ruhr-Universit\"at Bochum, 44780 Bochum, Germany}

\author[00000-0000-0000-0000]{Ilja Jaroschewski}
\affiliation{Theoretical Physics IV: Plasma-Astroparticle Physics, Faculty for Physics \& Astronomy, Ruhr University Bochum, 44780 Bochum, Germany}
\affiliation{Ruhr Astroparticle And Plasma Physics Center (RAPP Center), Ruhr-Universit\"at Bochum, 44780 Bochum, Germany}

\author[00000-0000-0000-0000]{Armin Ghorbanietemad}
\affiliation{Theoretical Physics IV: Plasma-Astroparticle Physics, Faculty for Physics \& Astronomy, Ruhr University Bochum, 44780 Bochum, Germany}
\affiliation{Ruhr Astroparticle And Plasma Physics Center (RAPP Center), Ruhr-Universit\"at Bochum, 44780 Bochum, Germany}
\author[0000-0001-5607-3637]{Imre Bartos}
\affiliation{Department of Physics, University of Florida, PO Box 118440, Gainesville, FL 32611-8440, USA}

\author[00000-0003-2769-3591]{Emma Kun}
\affiliation{Theoretical Physics IV: Plasma-Astroparticle Physics, Faculty for Physics \& Astronomy, Ruhr University Bochum, 44780 Bochum, Germany}
\affiliation{Astronomical Institute, Faculty for Physics \& Astronomy, Ruhr University Bochum, 44780 Bochum, Germany}
\affiliation{Ruhr Astroparticle And Plasma Physics Center (RAPP Center), Ruhr-Universit\"at Bochum, 44780 Bochum, Germany}
\affiliation{Konkoly Observatory, ELKH Research Centre for Astronomy and Earth Sciences, Konkoly Thege Miklós \'ut 15-17, H-1121 Budapest, Hungary}
\affiliation{CSFK, MTA Centre of Excellence, Konkoly Thege Miklós \'ut 15-17, H-1121 Budapest, Hungary}

\author[0000-0003-3948-6143]{Peter L.~Biermann}
\affiliation{MPI for Radioastronomy, 53121 Bonn, Germany}
\affiliation{Department of Physics \& Astronomy, University of Alabama, Tuscaloosa, AL 35487, USA}
\affiliation{Department of Physics \& Astronomy, University of Bonn, 53115 Bonn, Germany}




\begin{abstract}

On September 18, 2022, an alert by the IceCube Collaboration indicated that a $\sim 170$~TeV neutrino arrived in directional coincidence with the blazar TXS~0506+056. This event adds to two previous pieces of evidence that TXS~0506+056 is a neutrino emitter, i.e.\ a neutrino alert from its direction on September 22, 2017, and a $3\sigma$ signature of a dozen neutrinos in 2014/2015. 
\cite{deBruijn2020} showed that two previous neutrino emission episodes from this blazar could be due to a supermassive binary black hole (SMBBH) central engine where jet precession close to the final coalescence of the binary results in periodic emission. 
This model predicted a new emission episode consistent with the September 18, 2022 neutrino observation by IceCube. Here, we show that the neutrino cadence of TXS~0506+056 is consistent with a SMBBH origin.
We find that the emission episodes are consistent with a SMBBH with  mass ratios $q\lesssim 0.3$ for a total black hole mass of $M_{\rm tot}\gtrsim 3\cdot 10^{8}$~$M_{\odot}$.  For the first time, we calculate the characteristic strain of the gravitational wave emission of the binary, and show that the merger could be detectable by LISA for black hole masses $<5\cdot 10^{8}\,M_{\odot}$ if the mass ratios are in the range $0.1 \lesssim q \lesssim 0.3$.
We predict that there can be a neutrino flare existing in the still to be analyzed IceCube data peaking some time between August 2019 and January 2021 if a precessing jet is responsible for all three detected emission episodes. The next flare is expected to peak in the period January 2023 to August 2026. Further observation will make it possible to constrain the mass ratio as a function of the total mass of the black hole more precisely and would open the window toward the preparation of the detection of SMBBH mergers. 
\end{abstract}
\keywords{galaxies: active, neutrinos: galaxies, galaxies: individual (TXS~0506+056)}

\section{Introduction}
\label{section:intro}
The blazar TXS~0506+056 is of central interest in multimessenger astronomy, since several tantalizing hints of neutrino emission from this source have been published. First evidence at the $3\sigma$ level was seen when a neutrino of $\sim 300$~TeV energy was detected in coincidence with the direction of TXS~0506+056, while an intense flare at GeV gamma-ray energies shown by the same blazar was ongoing, as detected by Fermi-LAT \citep{ICTXS2018a}. The past 10 years of IceCube data from the direction of TXS~0506+056 were analyzed in a blind fashion subsequently, and another significant piece of evidence ($3.5\sigma$) was detected at the turn of the year 2014/2015. This potential flare was only found in an offline-analysis, as it is very different in its nature as compared to the flare from September 2017: the background deviation comes from an excess of $\sim 10$ events at $\sim 10$~TeV  \citep[][]{ICTXS2018b}. 
At the same time of the neutrino flare, the gamma-ray light curve is in a low state. 
The picture of a low gamma-state together with a high neutrino state is puzzling to begin with, as neutrinos and gamma-rays are co-produced. It can be best explained by gamma-absorption as outlined later. First evidence that neutrino sources must be connected to gamma-absorption was already seen in the first signal of the diffuse neutrino flux. Even here, theoretical models are in need of gamma-ray absorption in order not to overshoot the diffuse gamma-ray flux as measured by Fermi \citep{ahlers2015}. 
Further coincidences of high-energy neutrinos with blazars also point to the fact that these arrive at times of low gamma-ray activity \citep{Kun2020}. Even the IceCube event IC170922A, detected from the direction of TXS~0506+056 during a gamma-ray flare arrived at a time in which the gamma-ray emission was in a local minimum. All of these pieces of evidence point toward a scenario in which the production of high-energy neutrinos happens in very dense regions in which gamma-rays are absorbed and cascade down to MeV energies \citep{Halzen2020}.

The detection of a high-energy neutrino on September 18, 2022 represents the newest piece of evidence that TXS~0506+056 is in fact a neutrino emitter. The event is detected with an estimated energy of $\sim 170$~TeV and classified as a "bronze alert", with a signalness of 42\% \citep{ic220918A_gcn1}. The event is centered at the direction $(\mathrm{RA, Dec}) = (+05h 00m 36s, +03d 34' 48")$ (J2000 coordinates), with an uncertainty of $\sim 3.58^{\circ}$ (90\% containment). While typical track-like events have a much smaller uncertainty \citep{icecube_10yr_2020}, this event was skimming the edge of the detector and the track is therefore not fully contained, as described in \cite{ic220918A_gcn1}. Follow-up observations by the optical MASTER-Amur robotic telescope resulted in upper limits up to a magnitude of 18.9 \citep{IC220918A_optical_master}. In a search of IceCube data of $1000$~seconds and 2 days centered around the arrival time of IC220918A, no further track was found   
\citep{IC220918A_gcn2}. An analysis of Fermi data reveals seven Fermi-detected sources in the 90\% uncertainty interval of the event, among which the source Fermi J0502.5+0037 was newly detected in this dedicated search in 14 years of Fermi data \citep{IC220918A_fermi}.

With this new event, multimessenger modeling is confronted with a third potential neutrino flare from TXS~0506+056, again somewhat different from the one in 2017, as the gamma-ray light curve is in a low state. As pointed out by \cite{reimer2019}, modeling the two first flares in 2014/2015 and 2017 with the same emission model is difficult to impossible, while fitting the multimessenger data for the event IC170922A works quite well in the standard approach of high-energy cosmic-rays in a blob propagating along a jet axis, interacting with ambient photons and gas targets \citep{gao2019,rodrigues2019,petropoulou2020}. As the two flares are separated in time by about
 $\sim 2.5$~years, assuming that the plasmoid propagates rela\-tivistically along a jet axis, it is clear that the local environment in which cosmic rays interact with ambient targets can have changed significantly, so that even the flare properties can change with time.
 
 In \cite{deBruijn2020}, the hypothesis of TXS~0506+056 harboring a precessing AGN jet was made and future high-energy neutrino flares were predicted. The next flare was expected around 2019-2020, the next-to-next flare around 2022-2023. It was already argued in \cite{deBruijn2020} that the 2019-2020 flare could still hide in the offline data of the IceCube Neutrino Observatory. The new event IC220918A falls right into the period of a potential fourth flare and therefore strengthens the hypothesis presented in \cite{deBruijn2020}.
 
 In this paper, we present an update and extension of the model from \cite{deBruijn2020}, which allows us to substantiate our predictions concerning the prediction of the flaring periods, and gravitational wave signatures. We perform calculations at 2.5 post-Newtownian order of the flaring behavior of a jet that is precessing due to a supermassive binary black hole (SMBBH) in the center of the active core of the galaxy. In Section \ref{sec:predictFlare}, we present the general scheme of the model. The results including the prediction of the timing of a third neutrino flare, prediction of the emission of the merger of the SMBBH, as well as the expected signatures of gravitational waves of the merger are presented in Section \ref{sec:gwrad}. We close with a short summary of our findings and predictions.

\section{Spin--orbit precession  model}
\label{sec:predictFlare}

Massive galaxies typically host supermassive black holes, which can be formed through several mergers of smaller black holes over time \citep{Press1974, Conselice2003, Caramete2010}. 
It is therefore expected, that a larger fraction of galaxies hosts supermassive binary black holes (SMBBHs) \citep[e.g.][]{Volonteri2003}.
Evidence for such SMBBH systems can be found by looking for quasi-periodic emission from jets of active galaxies, as the binary nature of the central black holes lead to the prediction of jet precession \citep{GerPLB2009}. We apply this model of a precessing jet to the case of TXS~0506+056 and shortly summarize the model here. We base our calculations in the model presented in \cite{kun2022}, which is an extension of the model presented in
\cite{deBruijn2020}. Here, it is assumed that there is a jet that is oriented along the dominant spin of the SMBBH, which is given by an angle $\phi$ as a function of the remaining time until the binary coalescence, $\Delta T_{\rm GW}$. The angle $\phi$ covers the range $0\degr$ to $360\degr$ in one spin-orbit precession period and is defined to be in the plane perpendicular to the total angular momentum.  We slightly modified the model of \cite{deBruijn2020} (which in turn was based on earlier work by \cite{GerPLB2009}) as described in \cite{kun2022}. The resulting description of the angle $\phi$ as a function of $\Delta T_{\rm GW}$ and the mass ratio of the two black holes with the heavier mass $m_1$ and the lighter mass $m_2$, $q=m_2/m_1$, is given as
\begin{eqnarray}
    &&\phi(\Delta T_{\rm GW} \,,\, q) = - \frac{2 \, (4 + 3q)}{(1 + q)^2}\times \nonumber \\
           &&\, \times \left( \frac{5 \, c}{32 \,G^{1/3} M^{1/3}} \cdot \frac{(1 + q)^2}{q} \right)^{3/4} \left(\Delta T_{\rm GW}\right)^{1/4} 
           + \psi \,.
           \label{eq:JetModel}
\end{eqnarray}
Here, $G$ is the gravitational constant, $c$ is the speed of light and $M=m_1+m_2$ is the total mass of the SMBBH. 
Further, $\psi$ is an integration constant, which gives the initial direction of the jet in the inspiral phase of the merger before it changes significantly due to spin--orbit interactions. 

We apply the above model to the neutrino data by assuming that the flares from 2014/2015, 09/2017, and 09/2022 come from a jet precession motion.  We assume that the jet made a full rotation from the detection of the first to the second flare (2014/2015 to 09/2017). The connection between the two flares is therefore given as
\begin{equation}
    \phi(\Delta T_{\rm GW} \,,\, q) = \phi(\Delta T_{\rm GW} - P_{\rm jet} \,,\, q) \pm \zeta \,.
    \label{eq:determine_T_GW}
\end{equation}
Here, $P_{\rm jet}$ is the precession period.
The parameter $\zeta$ is introduced to model the half-opening angle of the jet in the equations, which translates into the duration of the flare in terms of observables.
The constant $\psi$ in Eq.\ (\ref{eq:determine_T_GW}) can be eliminated by inserting Eq.\ (\ref{eq:JetModel}), once as a function of $\Delta T_{\rm GW}$ and once as a function of $\Delta T_{\rm GW}-P_{\rm jet}$.

The above equations can now be used to  determine the further evolution of the systems, thus also predicting future flares. In our case, we actually do know the occurrence of the fourth flare, consistent with the prediction. This helps us to further pinpoint the possible location of the third flare as described in the next section. 
 
The results are being determined as a function of the mass ratio $q$, which can be constrained better the more flares have been detected.

\section{Signal prediction from TXS~0506+056}
\label{sec:gwrad}

\begin{figure*}
    \centering
    \includegraphics[angle=0,scale=0.5]{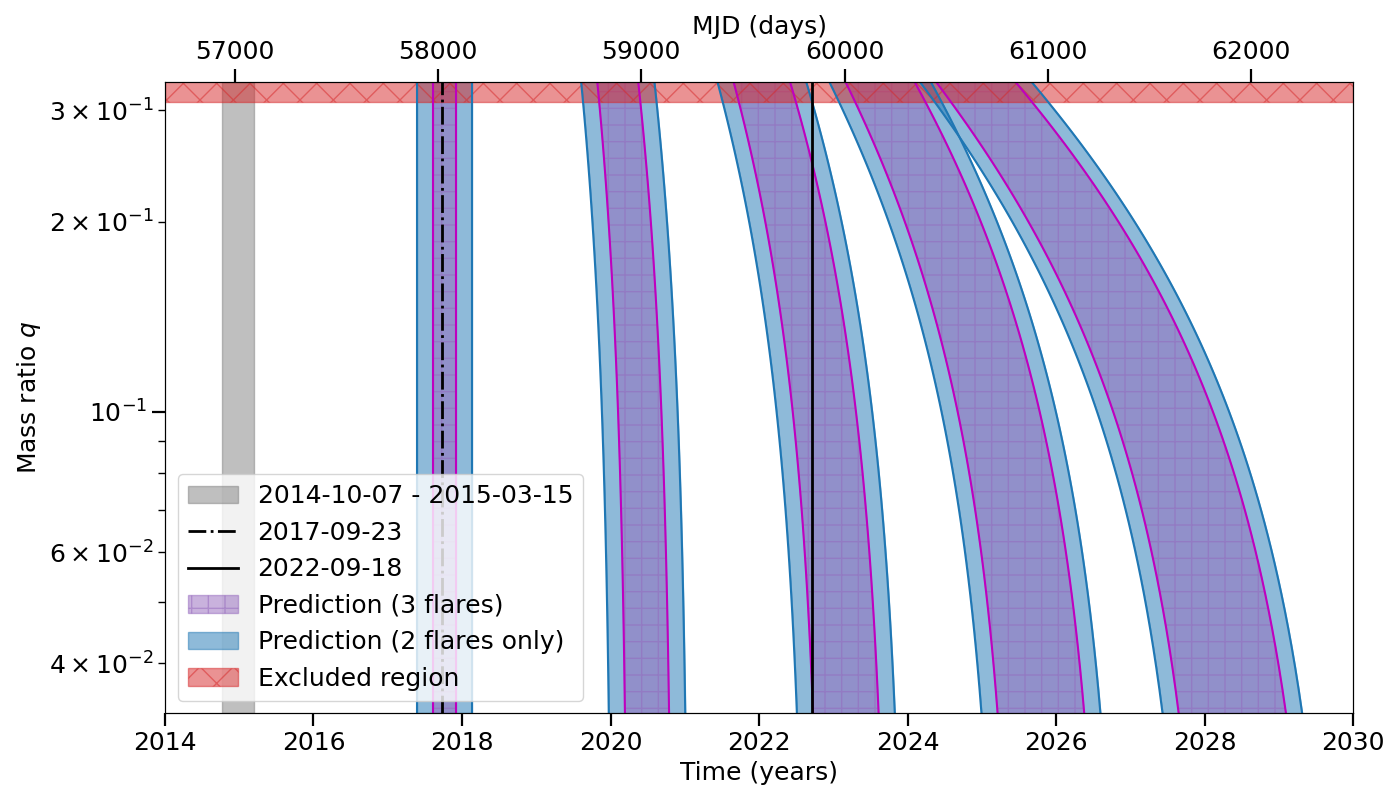}
    \caption{ Prediction of the time of the neutrino flares from TXS~0506+056 in 2.5 post-Newtonian order. Mass ratios of $q>0.3$ can be excluded by the detection of the three flares assuming a black hole mass of $M_{\rm tot}=3\cdot 10^{8}\,M_{\odot}$. The blue shaded regions show the predictions using the two first flares (gray shaded band and dot-dashed line) as a starting condition, the purple, hashed regions show the prediction when including the event from September 18, 2022 (black, solid line). The exclusion region of mass ratios $q>0.3$ is derived from the occurrence of the September 18, 2022 event.
    The blue and purple shaded regions around the second flare (September 23, 2017) show its possible time windows assumed in this model of two respectively three flares.
    }
    \label{fig:flarePredict}
\end{figure*}

Figure~\ref{fig:flarePredict} shows the flare predictions in dependence of the mass ratio $q$. The blue shaded areas are predictions that are using the first two flares (2014/2015, grey area and 09/2017, dot-dashed line) only. The event on September 18, 2022 (black, solid line) falls right into the prediction of the fourth flare. Including this as a known parameter, the predictions can be specified (purple, hashed regions). In particular, the detection of the flare constrains the mass ratio of the system to $q<0.3$, consistent with the range of typical mass ratios in the merger of $0.03-0.3$, see e.g.\ \citep{GerPLB2009}. Changing the total black hole mass that was assumed to be $M_{\rm tot}=3\cdot 10^{8}\,M_{\odot}$ here changes this limit somewhat, i.e.\ to $q<0.2$ for $M_{\rm tot}=10^{9}\,M_{\odot}$.
Within our model, we predict that any time between August 2019 and January 2021 a neutrino flare could exist in the still to be analyzed IceCube data. We also predict the occurrence of the next flare, which should peak any time between January 2023 and August 2026. The exact time of the flare will further constrain the mass ratio of the system as a function of the total mass.

We can further model the expected characteristic strain $h_c$, following \cite{Sesana2016}:
\begin{equation}
    h_c = \sqrt{\frac{2}{3}}
    \, \frac{1}{r(z)}
    \, \frac{1}{\pi^{2/3} c^{3/2} } 
    \, \frac{\left(G \mathcal{M}\right)^{5/6}}{(1 + z)^{1/2}}
    \, f^{-1/6} \,,
\end{equation}
Here,  $r(z)$ is the comoving source distance, $\mathcal{M} = (m_1 m_2)^{3/5}/(m_1 + m_2)^{1/5} $ is the chirp mass, and $f$ the observed GW frequency.

\begin{figure*}
    \centering
    \includegraphics[angle=0,scale=0.5]{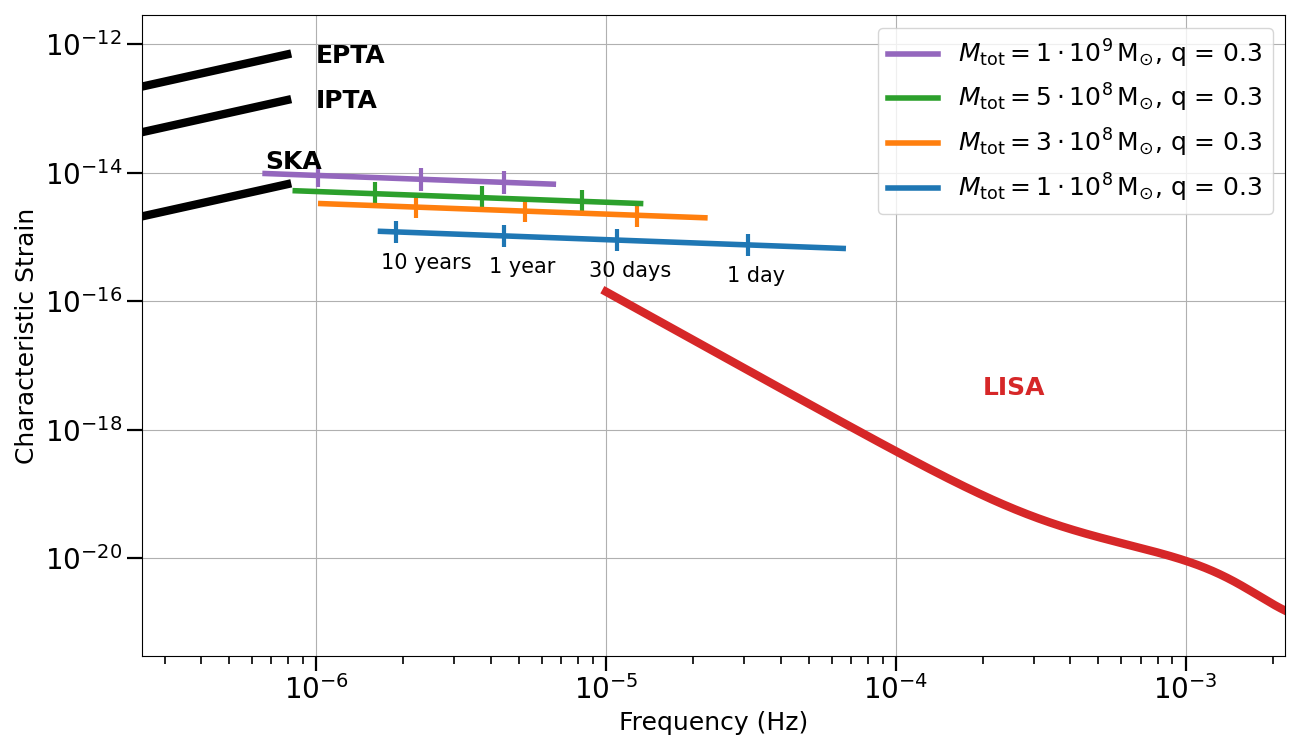}
    \caption{Characteristic strain expected for the merger of the SMBBH in TXS~0506+056 for a mass ratio of $q=0.3$ and for different total masses. The lowest curve (blue) is for $M_{\rm tot}=10^{8}\,M_{\odot}$, followed by orange ($M_{\rm tot}=3\cdot 10^{8}\,M_{\odot}$) and green ($M_{\rm tot}=5\cdot 10^{8}\,M_{\odot}$). The upper line (purple) shows $M_{\rm tot}=10^{9}\,M_{\odot}$. The dashes show the time to merger (10, 1 year, 30 days and 1 day, from the left. Except the lowest curve with $M_{\rm tot}=10^{8}\,M_{\odot}$, the 10 year dash is missing from the other curves as the time-to-merger as of today for the corresponding total masses and mass ratios is lower.
   }
   \label{fig:strain}
\end{figure*}

Figure~\ref{fig:strain} shows the expected characteristic strain for TXS~0506+056 for different total black hole masses and mass ratios. We show the figure for the maximum allowed mass ratio $q=0.3$. For lower mass ratios $q<0.3$, the intensity decreases and the strain at a fixed time to merger is shifted to lower frequencies. Four cases are shown in the figure from the top one: $M_{\rm tot}=10^{9},\,5\cdot 10^{8},\,3\cdot 10^{8},\, 10^{8}\,M_{\odot}$ (colors: purple, green, orange, and blue, respectively). Each line shows the characteristic strain expected as a function of the time to merger for the time span from the time-to-merger as of today (smallest frequencies) to the happening of the merger (largest frequencies). 
The current and future expected sensitivity curves from three pulsar timing arrays, the European Pulsar Timing Array (EPTA, 5 pulsars with 10 years of observation time), International Pulsar Timing Array (IPTA, 20 pulsars with 15 years of observation time), and Square Kilometer Array (SKA, 100 pulsars with 20 years of observation time), are shown in black
\citep{Moore2015}. The LISA sensitivity curve is shown in red \citep{Robson2019}.
Most interesting for SKA is the scenario of a heavy mass system, while LISA has easier access to the lighter black holes, simply due to the different frequency ranges. It can be seen that it will be difficult for SKA to observe this specific source, since it is too close to the final coalescence. LISA, on the other hand, has good access to black holes of masses below $5\cdot 10^{8}$~$M_{\odot}$. Especially for lower masses of $10^{8}$~$M_{\odot}$, the characteristic strains very close to the merger may happen during the uptime of LISA. So, in general what needs to happen is that the characteristic strain is at frequencies and intensities accessible to LISA, but also that it happens at the right time. Analyzing this parameter space, we can derive that for masses $M_{\rm tot} < 5 \cdot 10^8\,M_{\odot}$ and for mass ratios $0.1 \lesssim q \lesssim 0.3$.   So, if the neutrino emission is confirmed to be periodic by future data, TXS~0506+056 is a clear candidate for a detection of the merger in gravitational waves.

\section{Summary, Conclusions, and Outlook}
\label{sec:sumconcl}
In this paper, we have shown that the high-energy neutrino detected by IceCube Neutrino Observatory in spatial coincidence with TXS~0506+056 on September 18, 2022, was predicted by \cite{deBruijn2020}. With this new event, we can now sharpen the prediction of the timing of subsequent neutrino flares. We conclude that the mass ratio of the two black holes must fulfill $q<0.3$ for masses $M_{\rm tot}>3\cdot 10^{8}\,M_{\odot}$.
Within our model, we predict the following:
\begin{itemize}
    \item \textbf{Neutrino flare in the unanalyzed IceCube data:} a flare should be existing in the still to be analyzed IceCube data with the peak emission happening any time during the period August 2019 and January 2021
    \item \textbf{Upcoming neutrino flare during the lifetime of IceCube:} the next flare should peak in the time period January 2023 and August 2026. The exact time of the flare will further constrain the mass ratio of the system as a function of the total mass. 
    \item \textbf{Possible detection of gravitational waves:} In the uptime of LISA, assumed between 2034-2044, the parameter space allows gravitational wave detection for masses $M_{\rm tot} < 5 \cdot 10^8\,M_{\odot}$ and mass ratios of $0.1 \lesssim q \lesssim 0.3$.
\end{itemize}
The first flare that happened at the turn of the year 2014/2015 was also not captured by an alert, but remained hidden in the offline data until unblinding. The reason was that the signal consisted of a larger number of $\sim 10$ events above the expected atmospheric background, which all had a relatively moderate energy ($\sim 10$~TeV) so that individual events did not trigger alerts. The difference in the signatures, together with a multiwavelength behavior that is not periodic is the largest challenge for this model.
If we assume that the difference in the flaring behavior is of intrinsic nature, what might be considered is the existence of two jets instead of one. A binary system should have two jet systems in general, but typically with one of them dominating the system \citep{GerPLB2009}. \cite{Britzen2019} argue that there is evidence for a binary jet in the data. From a typical merger history, it is expected that the two jets can have similar precessing periods. If the jets are somewhat different in their intrinsic nature concerning parameters that determine the observables like the opening angle of the jet and the acceleration power, gas and photon field distribution can be very different, which would explain why the nature of the 2014/2015 flare and the hidden flare in 2020 are not detectable in the IceCube alerts, while the 2017 and 2022 events are caught by the alerts. The prediction would be that one jet produces a high-intensity signal, but with an energy cutoff at $\sim 100$~TeV proton energy (2014/2015 and 2020 flares), while the other one produces a flatter spectrum with a higher energy cutoff at $\sim $~PeV energies. But depending on the reason for the different behavior of the different flares, it is not clear yet what the 2019/2020 and 2023-2026 flares would look like exactly. 

As for the  gamma-ray emission, we argue that it is not correlated to the neutrino emission, as the neutrinos are produced in a gamma-absorbed environment. There is evidence that even other potential neutrino sources represent a dense environment in which gamma-rays are highly absorbed (see e.g.\ \cite{Kun2020,eichmann202}). These findings are in accordance with the fact that even the diffuse neutrino flux is too bright to come from transparent gamma-ray sources, because all models will overshoot the diffuse gamma-ray background for the observed spectral index. 

Finally,
the possible detection of the merger by LISA opens the exciting possibility of following up on a periodic neutrino source until the merger is detectable in gravitational waves years to decades later. This way, we are starting to finally identify binary black hole mergers early-on and to understand their physics by connecting multimessenger data of gamma-rays, neutrinos, and gravitational waves.

In order to establish this model of a precessing jet, more data especially from IceCube are needed. Once a periodicity can be confirmed in neutrinos, a theoretical model including all wavelengths can be developed. For now, this paper makes very precise predictions of what to expect from TXS~0506+056 in the near future, which makes this a model that can be tested on short timescales and adds to the fact that this is an exciting time for multimessenger astronomy.

\bibliography{fermi_neutrino}
\bibliographystyle{aasjournal}

\begin{acknowledgments}
We would like to thank Francis Halzen, Anna Franckowiak, Erin O'Sullivan, Wolfgang Rhode, and Marcos Santander for valuable comments and discussions.
 J.T., I.J., A.G., and E.K.\ acknowledge support from the German Science Foundation DFG, via the Collaborative Research Center \textit{SFB1491: Cosmic Interacting Matters - from Source to Signal} (project number 445052434) and from the project \textit{MICRO} (project number 445990517).
 I.B. acknowledges the support of the Alfred P. Sloan Foundation and NSF grants PHY-1911796, PHY-2110060 and PHY-2207661. E.K.\ thanks the Hungarian Academy of Sciences for its Premium Postdoctoral Scholarship.
\end{acknowledgments}

\end{document}